\newcommand{\bd}{\begin{document}}
\newcommand{\ed}{\end{document}}
\newcommand{\bqu}{\begin{quote}}
\newcommand{\equ}{\end{quote}}
\newcommand{\mb}{\makebox}
\newcommand{\hs}{\hspace}
\newcommand{\vs}{\vspace}
\newcommand{\hf}{\hspace*{\fill}}
\newcommand{\sth}{\,:\,}
\newcommand{\spot}[5]{\put(#1,#2){\makebox(0,0)[{#3}{#4}]{#5}}}
\newcommand{\spots}[8]{\multiput(#1,#2)(#3,#4){#5}%
{\makebox(0,0)[{#6}{#7}]{#8}}}
\newcommand{\bp}{\begin{picture}}
\newcommand{\ep}{\end{picture}}
\newcommand{\Vector}[5]{\put(#1,#2){\vector(#3,#4){#5}}}
\newcommand{\Vectors}[8]{\multiput(#1,#2)(#3,#4){#5}%
{\vector(#6,#7){#8}}}
\newcommand{\uprarw}[3]{\put(#1,#2){\line(-1,0){1.5}}
\put(#1,#2){\line(0,-1){1.5}}
\put(#1,#2){\line(-2,-1){#3}}}
\documentstyle[12pt]{article}
\begin{document}
\baselineskip24pt
\vfill
\centerline{\Large Coupled Maps on Trees}
\vspace{10pt}
\centerline {\bf Prashant M. Gade and Hilda A. Cerdeira}
\vspace{3pt}
\begin{center}
{\it International Centre for Theoretical Physics\\ P.O. Box 586, Trieste
34100, ITALY}\\
\vskip .5cm
and\\ {\bf Ramakrishna Ramaswamy}\\
\vspace{3pt}
{\it School of Physical Sciences\\ Jawaharlal Nehru University, New Delhi
110 067, INDIA}\\
\end{center}
\vspace{6pt}
\begin{abstract}
We study coupled maps on a Cayley tree, with local (nearest-neighbor)
interactions, and with a variety of boundary conditions.  The homogeneous
state (where every lattice site has the same value) and the
node-synchronized state (where sites of a given generation have the same
value) are both shown to occur for particular values of the parameters and
coupling constants.  We study the stability of these states and their
domains of attraction.  As the number of sites that become synchronized is
much higher compared to that on a regular lattice, control is easier to
effect.  A general procedure is given to deduce the eigenvalue spectrum
for these states.  Perturbations of the synchronized state lead to
different spatio-temporal structures.  We find that a mean-field like
treatment is valid on this (effectively infinite dimensional) lattice.
\end{abstract}
\vspace{6pt}
PACS number(s):05.45. +b, 47.20. Ky
\pagebreak
\section{Introduction:}
Coupled map lattices (CML) have been explored in a variety of contexts in
recent years, particularly as prototypes of spatially extended
systems.  These are simple models wherein both space and time play a role;
furthermore, it is anticipated that the insight gained over the past two
decades in studying low-dimensional nonlinear dynamical systems can be
profitably exploited in providing an understanding of such complex
systems\cite{book}.

The phenomenology displayed by coupled maps on regular one and two
dimensional lattices has been extensively studied by Kaneko\cite{KK}. In
addition, CML have been used to model a wide variety of complex phenomena,
such as the study of the kinetics of phase ordering processes\cite{OP},
crystal growth\cite{KLR}, neuronal systems \cite{Wan}, optical fibres
\cite{Per} and pattern formation\cite{Bark}. Chat\'e and
Manneville have also used CML to model spatiotemporal
intermittency\cite{CM}.  A route to a spatiotemporally inhomogeneous state
through wavelength doubling bifurcations has also been recently
identified\cite{GA}.

In this paper, we study coupled maps on a Cayley tree. This lattice is
embedded in infinite dimensions, and thus should give some indication of
CML phenomenology in higher dimensions\cite{inf}. Although the Cayley tree
(sometimes termed the Bethe lattice) is an idealized hierarchical lattice
with no immediate physical application, it is convenient for study since
there are no closed loops.  Furthermore, the Bethe lattice is the simplest
sort of branching media model encountered in many physical processes.

Previous studies of coupled map systems have largely (except for a study
by Cosenza and Kapral\cite{Kap} of CML on a Sierpinski gasket) been
carried  on with local coupling on regular lattices in 1- and 2-
dimensions, or with global coupling, in which case there is no notion of
lattice geometry.  Our motivation in choosing the Bethe lattice is
two-fold. Apart from the mathematical convenience, it is worth considering
that in many physical situations, the medium supporting dynamics could be
nonuniform; in cases like chemical reactions in porous media or on
diffusion-limited-aggregation clusters, heterogeneity can lead to a
hierarchical structures\cite{RK}.  We note that hierarchical
structures have long since been studied also in spatiotemporal systems like
neural nets also due to their exponentially higher storage capacity\cite{Amit}.

A related question of some current interest is the control of
macroscopically cascaded dynamical systems. The synchronization of a large
set of oscillators connected in series\cite{AGS,Auerbach,exp,exp1} in a
given geometry and with particular boundary conditions, is directly
related to the problem of whether a similar CML can support a synchronized
state\cite{fn1}. We address this problem, and show below that the
criterion in CML for a synchronised (but chaotic) state to be stable is
that only {\it one} Lyapunov exponent is positive and all the rest are
negative. In both the cases, i.e. synchronization of the coupled oscillators
or  coupled maps the essence of the problem lies in the nature
of the eigenvalues and eigenvectors of the interaction matrix. In the
present work we deal with the situation of asymmetric coupling which is
easily obtained in experiments\cite{expas}.

The plan of this article is as follows. We define our model, and the
boundary conditions, and show that the stability of synchronized states
depends on the spectrum of eigenvalues of the interaction matrix. This is
related to the connectivity matrix for the lattice, and has a
singular-continuous structure.  Different patterns can arise from the
secondary instabilities. We note that for coupled piecewise linear maps
the study of the interaction matrix gives the whole Lyapunov spectrum,
which is related to the correlation length.

In several recent studies of globally coupled maps\cite{KK,sinha} a
breakdown of the ``law of large numbers" has been observed.  We find that
for this system, even with local coupling, the mean-field description is
valid in the macroscopic limit. This is probably related to the infinite
dimensional nature of the lattice.

\section{The Model:}
We study CML on a Cayley tree with coordination number 3, which we
construct in stages as follows. At the first level, there are three
branches, each of which split in two and each of them further bifurcates
and so on.  Each site thus has one parent and two daughters, except for
the origin which has no parent site and 3 daughters, and boundary sites
which have no daughters.  Each site on the lattice is assigned a variable
$x$, which evolves according to a deterministic rule depending on its own
value and the value of the nearest neighbors.  The evolution rule is taken
to be the following:
\begin{equation}
x(i,t+1) = h_0f(x(i,t)) + h_pf(x(i_p,t)) + h_d \sum_{j \in i_d}f(x(j,t))
\end{equation}
where $f$ is the function that determines how the lattice variables
evolve. A common choice for $f$ is the logistic map, $f(x) = \mu x(1-x)$.
The notation above is as follows: $i_p$ is the parent of site $i$, while
$i_d$ are the daughters.  $h_0, h_p$ and $h_d$ are coupling constants and
are taken such that $h_0+h_p+2h_d = 1$. Thus the evolution is contained in
the same phase space as that of a single map, {\it {e.g.}}~ [0,1] in case of
logistic map.  We also assume that the couplings are positive though most
of our results do not explicitly require this.

The evolution rules for the origin and for boundary sites are somewhat
different, since the former does not have a parent and the latter do not
have daughters. For the origin,
\begin{equation}
x(0,t+1)= o_0f(x(0,t)) + o' \sum_{j \in 0_d} f(x(j,t))
\end{equation}
while for the boundary sites,
\begin{equation}
x(k,t+1) = b_0f(x(k,t))+ b'f(x(k_p,t))
\end{equation}
with different choices for $o_0, b_0, o', b'$ to ensure that the dynamics
remains in the same phase space.

Each point on the Bethe lattice can be indexed by a string $a=\{a_1, a_2,
\ldots,a_i\}_i$, where $i$ denotes generation.
$a_1$ can take the single value $a_1 = 0$, $a_2$ can be $0, 1 $ or $2$,
and for $j~>~2$, the $a_j$'s take two values, $a_j=0$~or~$1$. With this
notation, it is possible to assign a unique number $n(A) =
n(a_1,\ldots,a_i)$ to each lattice point.
\begin{equation}
n(a_1,a_2,\ldots,a_i)=g(i-1)+a_i +2a_{i-1}+2^2a_{i-2}+\ldots+ 2^{i-2}a_2+1
\label{ndef}
\end{equation}
where
\begin{equation}
g(i)=1+3(2^{i-1}-1)
\label{gdef}
\end{equation}
is the total number of sites at the $i$th generation and g(0)=0.  (See
Fig.~1). We can thus formally write
\begin{equation}
X(t+1)=I F(X(t))
\end{equation}
where $F(X(t))$ is a column matrix, $X(t)$ is an array of variable values
assigned to the lattice points arranged in ascending order of $n(A)$, and
$I$ is the interaction matrix.  For example, for the Cayley tree with 3
generations {\it {i.e.}}~10 sites, the equation above reduces to
\begin{equation}
X(t+1)=
\left(\begin{array}{lccccccccr}
o_0&o'&o'&o'&0&0&0&0&0&0\\ h_p&h_0&0&0&h_d&h_d&0&0&0&0\\
h_p&0&h_0&0&0&0&h_d&h_d&0&0\\ h_p&0&0&h_0&0&0&0&0&h_d&h_d\\
0&b'&0&0&b_0&0&0&0&0&0\\ 0&b'&0&0&0&b_0&0&0&0&0\\ 0&0&b'&0&0&0&b_0&0&0&0\\
0&0&b'&0&0&0&0&b_0&0&0\\ 0&0&0&b'&0&0&0&0&b_0&0\\ 0&0&0&b'&0&0&0&0&0&b_0
\end{array}\right)F(X(t))
\end{equation}
where
\begin{equation}
F(X(t))=(f(x(1,t)),f(x(2,t)),f(x(3,t)), \ldots,f(x(g(k),t))^T
\end{equation}
where $T$ denotes transpose.  Note that the interaction matrix $I$ is such
that the evolution can be written as
\begin{equation}
x(l,t+1)=\sum_j I_{lj}f(x(j,t))
\end{equation}
where the sites have been labeled by the unique number $n(A)$.

\section{Coherent Patterns and Their Stability:}
Inspection of the symmetries of the lattice allow for the determination of
allowed coherent patterns. It is easy to verify that if one starts with
the pattern in which all the points at  each generation $i$ are assigned
the same value $z_i(t)$ at time $t$, the nature of the pattern cannot
change in time since points at the same generation have equivalent
evolution rules and remain synchronized. With parameters
\begin{equation}
o_0+3o' = b_0+b' = h_0+h_p+2h_d = 1
\label{cdef}
\end{equation}
another simple pattern is possible. This is also node-homogeneous, with
$z_i=z$ for all $i$, {\it {i.e.}} all the points on the lattice are
synchronized
since the evolution is essentially that of a single map $f$.

These `allowed' patterns can be observed in practice if and only if they
are linearly and convectively stable against small perturbations. In the
present work, we have mainly dealt with linear stability analysis of this
system in the stationary frame, and while we have not analysed convective
stability, numerical
experiments suggest that no extra instabilities other than the ones in an
equivalent one dimensional model creep in.  This directly evolves on the
fact that there are no loops on the lattice-- there is only one direction
in which instabilities can be enhanced in a moving frame of reference, and
these are the same as in the equivalent 1-d model.

For the linear stability analysis the eigenvalues and eigenvectors of the
matrix $J=\lim_{\tau\to\infty} J(\tau)$ where $J(\tau)= J_\tau\ldots
J_2J_1$, are (asymptotically) relevant.  The Jacobian matrix at time
$t$, {\it {i.e.}}~ $J_t$
is given by $J_t(i,j)=I(i,j)f'(x_j(t))$ and $x_j(t)=x(t)$
for all $j$.  Thus the Jacobian matrix is $J=lim_{\tau\to\infty}[I]^\tau
f'(x_\tau)f'(x_{\tau-1})\ldots f'(x_1)$.  The eigenvalues of $J$ are
$\lim_{\tau\to\infty} \lambda_i^\tau$ where $\lambda_i =v_i
\lambda$ where $v_i,~i~ =~ 1,2,\ldots, g(k)$ are the eigenvalues
of the interaction matrix $I$ and $\lambda=\lim_{t\rightarrow\infty}
\vert f'(x(t))f'(x(t-1))\ldots f'(x(1))\vert^{1/t}$. The
relevant eigenvectors are those of $I$, and the problem reduces to a study
of the eigenvalues and eigenvectors of the interaction matrix.

The fact that coherent patterns are allowed (by the condition in
Eq.~\ref{cdef}) implies that a right eigenvector of the interaction matrix
is $e_1~=~[1,1,\ldots,1]$.  This is a characteristic of row stochastic
matrices, and corresponds to the eigenvalue $\lambda$ for the product of
the $J$'s.  From Greshgorin's theorem\cite{Barnett} this is the largest
eigenvalue.  Consider a small deviation, $\Delta_0=[\delta_1,
\delta_2, \ldots,\delta_{g(k)}]$ from the homogeneous pattern
$[x,x,\ldots]$. We can reexpress $\Delta_0$ in the basis of eigenvectors
$e_1,e_2,\ldots,e_{g(k)}$
\begin{equation}
\Delta_0=a_1e_1+a_2e_2+\ldots +a_{g(k)}e_{g(k)}.
\end{equation}
After $t$ iterations the deviation from the homogeneous condition will be
\begin{equation}
\Delta_t=a_1\lambda_1^te_1+a_2\lambda_2^te_2+\ldots+a_{g(k)}
\lambda_{g(k)}^t e_{g(k)}.
\end{equation}
If the only eigenvalue with modulus greater than unity is
$\lambda_1~=~\lambda $ and $\vert \lambda v_j \vert <1$ for $j>1$,
{\it {i.e.}}~ the rest are less than unity in magnitude, then for large
enough $t$ we can write
\begin{equation}
\Delta_t\simeq a_1\lambda_1^te_1
\end{equation}
The perturbation grows along the direction $e_1=[1,1,\ldots,1]$ and any
random deviation will eventually be homogenized.

Thus the necessary (though not sufficient)
condition for the synchronized pattern to exist (and
evolve chaotically in time) is that $\lambda_1$ is the only eigenvalue
greater than unity and all others are less than unity in magnitude: a
linearly stable coherent pattern-- in the infinite lattice limit--
therefore requires a finite {\it gap} in the eigenvalue spectrum of the
interaction matrix.

The interaction matrix is analogous to the tight-binding Hamiltonian on
the Bethe lattice\cite{Chen} although the eigenvectors are different
(since the matrix is not necessarily symmetric or hermitian). However,
using similar arguments\cite{Chen}, it can be seen that all the sites at a
given generation are equivalent in the sense that if sites at every
generation are synchronised, this pattern will continue to exist in
absence of small perturbations or noise, since the evolution rule is the
same for all of them. Furthermore, one can see that if any two sites which
have the same parent are interchanged along with their sub-trees, the
system is left unchanged.  Using the equivalence of all points at a
generation and the permutation symmetries of the lattice, the similarity
transformation that will block-diagonalize the interaction matrix can be
deduced to be
\begin{equation}
S= \left(\begin{array}{lccccccccr} 1&0&0&0 &0 &0 &0 &0&0&0\\ 0&1&0&1 &0 &1
&0&0&0&0\\ 0&1&0&1 &0 &-1&0 &0&0&0\\ 0&1&0&-2&0 &0 &0 &0&0&0\\ 0&0&1&0 &1
&0 &1 &1&0&0\\ 0&0&1&0 &1 &0 &1 &-1&0&0\\ 0&0&1&0 &1&0 &-1&0&1&0\\ 0&0&1&0
&1 &0 &-1&0&-1&0\\ 0&0&1&0 &-2&0 &0&0&0&1\\ 0&0&1&0 &-2&0 &0 &0&0&-1
\end{array}\right).
\label{sdef}
\end{equation}
(The first three vectors follow from the fact that lattice points at each
generation are equivalent. The fourth and sixth vectors simply represent
the two linearly independent and mutually orthogonal interchanges possible
between points at the second generation, ($(1,0,-1) \pm (0,1,-1)$) while
the fifth and seventh vectors are similar interchanges within siblings
with the phase derived from the parent site.  The last three vectors arise
from the interchange between the siblings of the same parent.)

Thus the block diagonalizing matrix is written using permutation
symmetries of the underlying lattice; the blocks are as follows.
\begin{equation}
\left(\begin{array}{lcr}
o_0&3o'&0\\ h_p&h_0&2h_d\\ 0&b'&b_0\\
\end{array}\right)
\end{equation}
The two doubly degenerate eigenvalues are the eigenvalues of the matrix
given below.  They correspond to the fact that one can have two
independent permutations in the three branches at the first node.
\begin{equation}
\left(\begin{array}{lr}
h_0&2h_d\\ 3b'&b_0\\
\end{array}\right)
\end{equation}
Finally we have a triply degenerate eigenvalue is $b_0$ (reflecting the
fact that one can have permutations between the daughters of any of the
three branches at the second node without affecting the matrix). One can
see that the consecutive blocks giving eigenvalues are just like earlier
blocks except that first row and column of the matrix are removed.  This
construction can be trivially extended to a matrix of higher order.  The
matrix $S^{-1}IS$ is block diagonal.

This scheme can be generalized to higher dimensions and the diagonalizing
matrix for the $k$th stage can be deduced as following. Specifying the
non-zero components of the column vectors (in the notation of
Eq.~\ref{ndef} to denote the components) the first $k$ vectors are as
follows,
\begin{equation}
\begin{array}{l}
v^1_{n(a_1)}=1,\\ v^2_{n(a_1,a_2)}=1,\\ v^3_{n(a_1,a_2,a_3)}=1,
\end{array}
\end{equation}
\ldots and
\begin{equation}
\begin{array}{l}
v^k_{n(a_1,a_2,a_3 \ldots ,a_k)}=1.
\end{array}
\end{equation}
({\it {e.g.}}~ the first 3 columns of the matrix defined in Eq. \ref{sdef})
Then we
have two sets of $k-1$ vectors. One is
\begin{equation}
\begin{array}{l}
\{ v^{k+1}_{n(0,1)}=v^{k+1}_{n(0,2)}=1, v^{k+1}_{n(0,3)}=-2 \},\\
\{v^{k+2}_{n(0,1,a_3)}= v^{k+2}_{n(0,2,a_3)}=1, v^{k+2}_{n(0,3,a_3)}=-2\},
\end{array}
\end{equation}
\ldots and
\begin{equation}
\begin{array}{l}
\{v^{k+(k-1)}_{n(0,1,a_3,...a_k)}=v^{k+(k-1)}_{n(0,2,a_3,...a_k)}=1,
v^{k+(k-1)}_{n(0,3,a_3,...a_k)}=-2\}.
\end{array}
\end{equation}
(See {\it {e.g.}}~the
fourth and fifth columns in Eq. \ref{sdef}.) The other set is
\begin{equation}
\begin{array}{l}
\{v^{2k}_{n(0,1)}=1, v^{2k}_{n(0,2)}=-1\},\\
\{v^{2k+1}_{n(0,1,a_3)}=1, v^{2k+1}_{n(0,2,a_3)}=-1\}
\end{array}
\end{equation}
\ldots and
\begin{equation}
\begin{array}{l}
\{v^{2k-1+(k-1)}_{n(0,1,a_3,\ldots,a_k)} = 1,
v^{2k-1+(k-1)}_{n(0,2,a_3,\ldots,a_k)}=-1\}.
\end{array}
\end{equation}
(See {\it {e.g.}}~ the
sixth and seventh columns in Eq. \ref{sdef}.) The next three
sets of $k-2$ vectors each are given as follows.  The first is of the type
\begin{equation}
\begin{array}{l}
\{v^{3k-1}_{n(0,1,1)}=1, v^{3k-1}_{n(0,1,2)}=-1\},\\
\{v^{3k}_{n(0,1,1,a_4)}=1,v^{3k}_{n(0,1,2,a_4)}=-1\},
\end{array}
\end{equation}
\ldots
and
\begin{equation}
\begin{array}{l}
\{v^{3k-2+(k-2)}_{n(0,1,1,a_4, .. ,a_k)}=1,
v^{3k-2+(k-2)}_{n(0,1,2,a_4, \ldots, a_k)}=-1\}.
\end{array}
\end{equation}
The second is
\begin{equation}
\begin{array}{l}
\{v^{4k-3}_{n(0,2,1)}=1, v^{4k-3}_{n(0,2,2)}=-1\},\\
\{v^{4k-2}_{n(0,2,1,a_4)}=1, v^{4k-2}_{n(0,2,2,a_4)}=-1\}
\end{array}
\end{equation}
\ldots and
\begin{equation}
\begin{array}{l}
\{v^{4k-4+(k-2)}_{n(0,1,1,a_4, .. ,a_k)}=1,
v^{4k-4+(k-2)}_{n(0,1,2,a_4, \ldots, a_k)}=-1\}.
\end{array}
\end{equation}
The last set is \[\{v^{5k-5}_{n(0,3,1)}=1, v^{5k-5}_{n(0,3,2)}=-1\}\] and
so on. (The last three columns in Eq.~\ref{sdef} are $v^{3k-1}$,
$v^{4k-3}$ and $v^{5k-5}$. For $k > 3$ newer sets will appear.) Now we
will have sets of vectors that will give blocks of size $k-3$.  The next
six blocks of $k-3$ vectors arise from permutations between the points on
the fourth generation and their descendants and are of the same type as
the three sets of $k-2$ vectors mentioned above which result from three
independent permutations possible between the six points on the third
generation.  One can continue this scheme untill reaching the boundary.
The number of
points on the boundary is $g(k)-g(k-1)=3~2^{k-2}$ and $(g(k)-g(k-1))/2 =
3~2^{k-3}$ permutation vectors are possible (see Eq.~\ref{gdef}.) which
will give a block of size 1 with the same degeneracy as the number of
permutations possible on the boundary.

For boundary conditions in which $o_0=h_0+h_p, o'=2h_d/3$, $b_0=h_0+2h_d,
b'=h_p$, at stage $k$, the first block of the block-diagonal form is
\begin{equation}
\left(\begin{array}{lccccr}
h_0+h_p &2h_d &0 &\ldots &0 &0\\ h_p &h_0 &2h_d& &0 &0\\
\vdots  &     &    &       & &\vdots\\
0 &0 &0 & &h_0 &2h_d\\ 0 &0 &0 &\ldots &h_p &h_0+2h_d\\
\end{array}\right)
\end{equation}
which exploits the equivalence symmetry of all the sites at a given
generation.  The second block, which exploits the permutation symmetry of
the points on the first generation is
\begin{equation}
\left(\begin{array}{lccccr}
h_0 &2h_d &0 &\ldots &0 &0\\ h_p &h_0 &2h_d & &0 &0\\
\vdots &       &       &   & &\vdots\\
0 &0 &0 & &h_0&2h_d\\ 0 &0 &0 &\ldots &h_p&h_0+2h_d\\
\end{array}\right)
\end{equation}
and so on. The last two blocks are
\begin{equation}
\left(\begin{array}{lr}
h_0&2h_d\\ h_p&h_0+2h_d\\
\end{array}\right)
,\left(\begin{array}{c} h_0+2h_d\\
\end{array}\right)
\end{equation}
The first block of order $k$ appears once in the block diagonal form, the
second of order $k-1$ appears twice, and next blocks of order $k-n$
($k-1\geq n\geq 2$)appears $3 \; 2^{n-2}$ times.  The first $k$
eigenvalues are therefore nondegenerate, then $k-1$ eigenvalues are doubly
degenerate, $k-n$ eigenvalues have degeneracy $3 \; 2^{n-2}$ for $k-1\geq
n \geq 2$ etc.

{}From the structure of the matrices and their degeneracies, one sees that
for the Cayley tree with one more generation, the $k-1$ degenerate blocks
are retained (with, however, the doubly degenerate block becoming triply
degenerate and other blocks doubling their degeneracy) with an additional
block which has degeneracy 2.  The block corresponding to non-degenerate
eigenvalues is, however, completely changed. The density of states has to
be singular continuous since the new eigenvalues that are created have a
lower degeneracy: the eigenvalue spectrum is a sum of delta peaks and is
nowhere differentiable, as is common in hierarchical systems\cite{Kap}.

In the situation when the synchronized state is linearly stable in
the stationary frame, the
typical degeneracy structure of eigenvalues is as shown in Fig.~2, for the
parameters and boundary conditions as discussed below.  The structure is
generic if the system is linearly stable, but the width of the gap
varies with the parameters.  For
piecewise linear maps, {\it {e.g.}} $ f(x) = rx~{\rm mod}~y$, the Jacobian is
constant in time, and the spectrum of eigenvalues of the interaction
matrix determines the Lyapunov spectrum of the CML, and thus (via the
Lyapunov dimension) the fractal dimension.

We can see from the degeneracy structure that about a quarter of the
eigenstates have their support fully from the boundary. The next layer is
approximately half of this number, and so on, with the number of states
which have their support up to a length $l$ from the boundary reducing
exponentially. This is in keeping with the expectation that the rate must
be faster than that in finite dimensional spaces where the number of modes
with wavenumber $\vert \kappa\vert <{\bf\kappa}$ is proportional to
${\bf\kappa}^d$.

Note that in the block-diagonal matrix, the blocks are tridiagonal and
(for positive couplings) all elements are positive. Such a matrix can be
transformed to symmetric form\cite{Barnett} and thus all its eigenvalues
are real: there can be no Hopf bifurcation leading to the instability of
a synchronized state.

Since the consecutive blocks are the principal (tridiagonal) submatrices
of the earlier block, the eigenvalues are interlaced\cite{Ostas}.  In
other words, the bounds for the eigenvalues of the lower block are
contained in the bounds for the eigenvalues for the higher block, and it
is enough to consider the first two blocks in order to study the stability
of spatially synchronized state.

For the first block, the non-degenerate eigenvalues are given by
$h_0+h_p+2h_d$ which is set to 1 by definition and the other $k-1$
eigenvalues are $h_0+2\sqrt{2h_dh_p}cos(\theta)$ where $\theta =2\pi
i/k,i=1,2,\ldots,k-1$.  The eigenvalues will have a gap if $2h_d\neq h_p$.

Consider the second block of order $m = k-1$, which is tridiagonal, and
can be symmetrized using a similarity transformation involving a diagonal
matrix with elements $D_{i,i}=[\sqrt{{2h_d}/h_p}]^{i-1}$.  This yields a
tridiagonal matrix $O$ such that the diagonal elements remain unchanged
and all the elements on upper and lower diagonal are $\sqrt{2h_dh_p}$.
Using Greshgorin's theorem\cite{Barnett} again, one can see that the
largest eigenvalue cannot exceed $h_0+2\sqrt{2h_ph_d}$ if $h_p>2h_d$.  As
explained above, the analysis of the first two blocks suffices to explore
the stability of synchronized state and thus the other blocks do not
modify the gap in the eigenvalue spectrum of the first block if $h_p > 2
h_d$.

Aranson, Golomb and Sompolinsky\cite{AGS} consider asymmetrically coupled
1-d chains with open boundary conditions where there is a convective
instability of synchronized patterns: perturbations {\it grow} in the
moving frame of reference, destroying macroscopic coherence. As we
have shown above, under these conditions macroscopic chaos is linearly
stable in stationary frame also on the Cayley tree.
However, the
difficulty in synchronizing large systems is less pronounced
in this case.  Due to the ultrametric topology, much larger systems can be
synchronized under the conditions above.  With open boundary conditions and
asymmetrical coupling, coherence is more easily established in the present
case.  For example, for $h_1 = 0.7$ in one direction and $h_2 = 0.1$ in
the other direction Aranson\cite{AGS} have a coherence length of around 55
for the choice of map $f(x)=a-x^2$ with a value of $a$ such that the
eigenvalue for a single map is 1.26.  Taking $ h_p=.7$, $2h_d=.1$, Fig. 2
shows a plot of eigenvalues as a function of degeneracies at these
parameter values for 50 generations. One can clearly see the gap between a
single nondegenerate eigenvalue above unity and the others below unity.
For the above parameters we can easily obtain a coherent pattern for $k =
20$ with random initial conditions: a CML with $\approx 10^6$ sites is
easily synchronized\cite{fn} to within $10^{-4}$ even under single
precision (16 binary digits) evolution! This is in stark contrast to 1-d
coupled CML, which has a coherence length of about 55 sites. This example
is a good illustration of the dramatic increase in stability with
hierarchical connectivity.

To check that no other instabilities than the ones expected from an
equivalent one-dimensional model come into picture, we looked at
the function $f(x)=1.39\;x \;mod \;1$ with the same
choice of coupling constants as above. Here the coherence is within
$10^{-5}$ for first 6 sites on a lattice of 1-dimensional lattice and even
on Cayley tree it is maintained for 6 generations. This is
expected, since there are no closed loops and the only direction in
which the instabilities can {\it flow and grow} is the one
from the centre to the boundary. However, we can see that since the
number of  sites synchronized is
exponentially higher on a tree with equivalent generations than an
one-dimensional lattice, exponentially
larger number of sites are synchronized on trees at equivalent parameters.
The base of exponent is related to the number of branches.

Auerbach\cite{Auerbach} has shown that one can circumvent the difficulty
arising from convective instabilities on a 1-d lattice by using
system-size dependent feedback control.  In essence, we achieve the same
ends through a change in geometry, without extra controls.  The boundary
conditions $o_0$ = $b_0$ = $h_0$, $o_p$ = $3b_d$ = $h_p+2h_s$ also gives
the same result, which indicates that some more variants are possible for
open boundary conditions and asymmetric coupling.

Now consider the node-homogeneous pattern. The stability matrix is given
by
\begin{equation}
J=\prod_{t=1}^{\infty} J_t
\end{equation}
Thus analysis of the eigenvalues of the product of matrices is reduced to
the analysis of the eigenvalues of the product of blocks. This is a great
simplification
since instead of considering matrices of order $2^k$, where $k$ is the
number of generations, we need consider $k$ matrices of order $k$ and
below. The analysis of the Jacobian matrix reduces to analysis of the
matrices
\begin{eqnarray}
\prod_{t=1}^{\infty}
\left(\begin{array}{lccccr}
o &3o'&0 &\ldots&0&0\\ h_p &h_0 &2h_d& &0&0\\ \vdots & & & & &\vdots\\ 0 &
0 & 0 & &h_0&2h_d\\ 0 &0 &0 &\ldots &b'&b_0\\
\end{array}\right)
\left(\begin{array}{l}
f'[z_1(t)]\\ f'[z_2(t)]\\ \vdots \\ f'[z_{k-1}(t)]\\ f'[z_k(t)]\\
\end{array}\right)
\nonumber\\,\prod_{t=1}^\infty \left(\begin{array}{lccccr}
h_0 &2h_d&\ldots &0&0\\ h_p&h_0&2h_d& &0&0\\ \vdots & & & & &\vdots\\ & 0
& 0 & &h_0&2h_d\\ &0 &0 &\ldots &b'&b_0\\
\end{array}\right)
\left(\begin{array}{l}
f'[z_2(t)]\\ f'[z_3(t)]\\ \vdots \\ f'[z_{k-1}(t)]\\ f'[z_k(t)]\\
\end{array}\right)
\end{eqnarray}
and so on.

Again the degeneracy structure is the same as for the interaction matrix,
and the Lyapunov spectrum is the sum of delta peaks and is a everywhere
discontinuous function, as for the fully synchronized state (which is a
special case of the node homogeneous structure).  We can similarly argue
that the condition for stability of the node homogeneous state (evolving
chaotically in time) is that the first block corresponding to the
nondegenerate eigenvectors is the only one with eigenvalues of modulus
greater than unity, all other blocks having eigenvalues with modulus less
than unity.  (This is since the first block corresponds to eigenvectors
that have contribution from all the generations, and the contribution from
all the points of the same generation is the same).

A simple example of such stable patterns can be constructed for $f(x)=
rx$~mod 1, with boundary conditions $o_0=0, b_0=0, o'=h_d, b'=h_p$, and
parameters $r=\sqrt(3)/2, h_0=0, h_d=h_p=1/3$.  For the Cayley tree with 5
generations {\it {i.e.}}~46 sites, it can be shown that the eigenvalues are
higher
than unity for the first block alone. Numerically, one can easily get node
homogeneous patterns, starting from random initial conditions. Thus the
possible coherent patterns are {\it characteristically} ~different from
those on regular lattices and the stability analysis is also
distinct\cite{unpublished}.

\section{The Infinite Dimensional Character:}
Now we study the properties of this model which should reflect the fact
that it is embedded in infinite dimensions, where a mean-field-like
treatment can be expected to be valid.  A collective
variable\cite{KK,sinha} $h(t)$ is defined as
\begin{equation}
h(t)={1 \over {g(k)}} \sum_{i=1}^{g(k)}x(i,t).
\end{equation}
where $g(k)$ is the total number of sites on the Cayley tree with $k$
generations as noted above, and $f(x)=\mu x(1-x)$ with $h_0=1-\epsilon$,
$h_p = h_d =\epsilon/3$, $b_0=o_0=h_0$, $b'=\epsilon$, $o'=\epsilon/3$
while the parameter values are $\epsilon = 0.1$ and $\mu=4$. The return
map of this variable
{\it {i.e.}}~$h(t+1)$ vs. $h(t)$  is a filled ellipse, whose size
decreases rapidly with the number of generations. We conjecture that in
the macroscopic limit it tends to a fixed point:
{\it {i.e.}}~though the evolution is chaotic for the system the
collective variable is invariant in time. The mean square deviation of
$h(t)$ decays like $1/N$ where $N$ is the number of sites (see Fig.~3),
quite unlike the case of globally coupled maps\cite{GCM}, where some
reorganization occurs in a way such that total number of independent
degrees of freedom is not linearly proportional to the number of sites.
This is not totally unexpected\cite{sinha} since the values being summed
are not independent random variables.  This also means that mean field is
not valid in these systems.  However, this expectation is fulfilled for
the Bethe lattice although the variables that are being summed are not
only not independent but are also not identically distributed: the
boundary evolves differently from the bulk and boundary effects are not
negligible in any limit since half the points reside at the boundary.
Fig.~4 shows the probability distribution of the central sites and the
boundary for the above case and they are clearly different.  However, the
sum behaves in a way that is expected from the sum of {\it iid} variables.
(We have verified that this behaviour holds at other values of the
parameters).

The recovery of mean field in infinite dimensions is interesting. Chat\'e
and Manneville \cite{Chate} have found that the mean field like approach
works better in higher dimensions in spatially extended systems and
connectivity plays a relatively marginal role. Further studies will be
necessary to determine the upper critical dimension for this problem.  We
are exploring this question.

\section{Conclusions:}

In summary, in this paper we have considered a coupled map lattice in an
ultrametric space. We show that synchronized but temporally chaotic
systems can be stabilized more easily in this space. We present a simple
method of obtaining the eigenvalue spectrum for the node homogeneous and
spatially synchronized structures using symmetries. We emphasize different
properties that owe their existence to the hierarchical connections and to
the infinite embedding dimension.

In an ecological model, Hogg {\it {et. al.}} \cite{Hogg} have suggested that a
hierarchically ordered random system should be more stable compared to
unstructured systems, and argue that the stability should scale like
$\log$ ({\it system size}) instead of ({\it system size}): thus
exponentially larger systems should be stabilized in hierarchical
organization.  Our results are in conformity with this expectation,
although the system studied in
\cite{Hogg} has a random branching. They have also found out that
assymetric interactions give a higher stability. This is also
expected from our analysis. It is difficult to find such
a naturally occurring system with a clearly demarcated tree-structure.
However, the properties we are trying to emphasize are qualitatively
unchanged so long as the connectivity remains free of loops (and the
resulting feedback).

Our results have immediate relevance to the problem of synchronization in
chaotic systems, which is currently evoking considerable
interest\cite{synch}. In 1-d, this problem has been extensively
discussed\cite{exp,exp1}.  Our results demonstrate a novel method of
stabilizing large systems and are of practical utility in this context.

\vspace{.03in}

{\bf Acknowledgements:}

We thank Deepak Dhar for helpful correspondence. RR would like to thank
the Condensed Matter Group of the ICTP for hospitality.

\vskip 1cm
\newpage
{\large Figure Captions}

\begin{itemize}

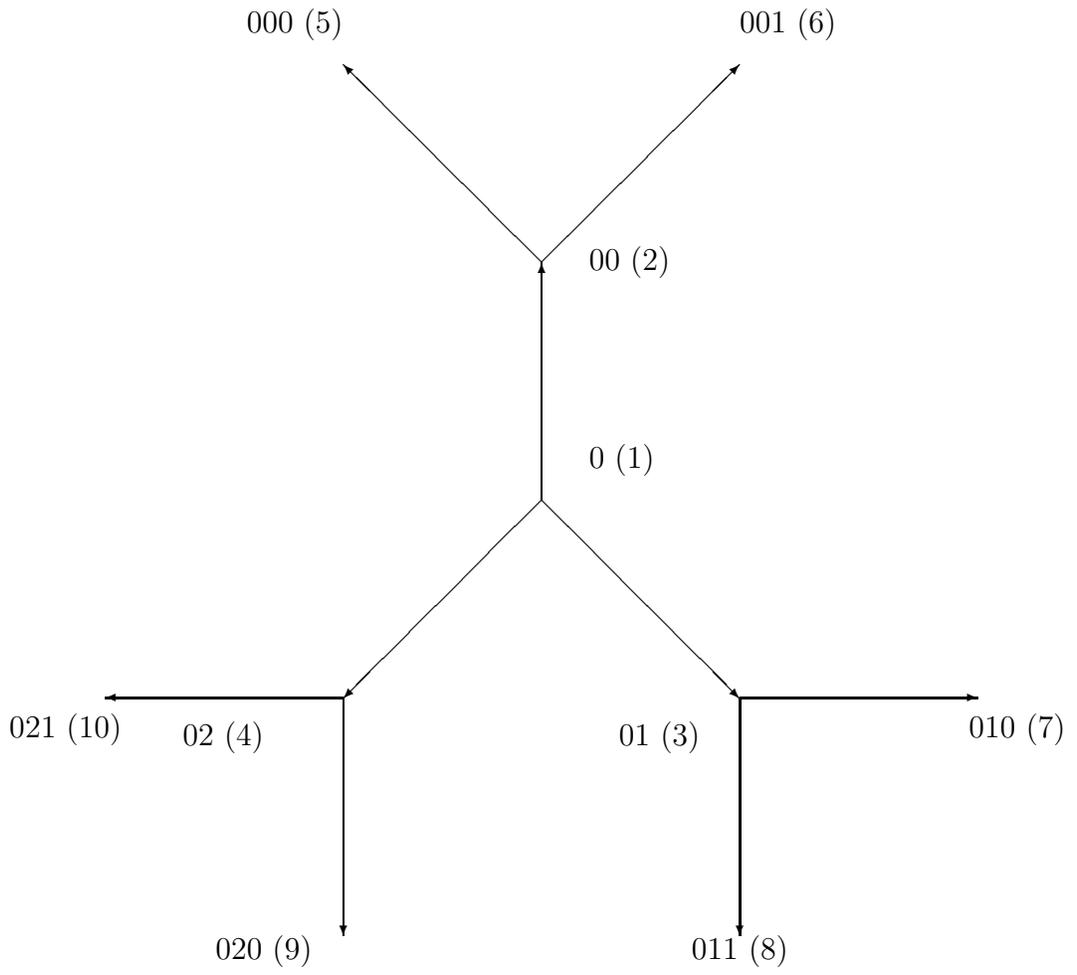
\item[Fig. 1]
The Cayley tree with three generations and the labeling scheme.
Arrows are directed towards the daughters.
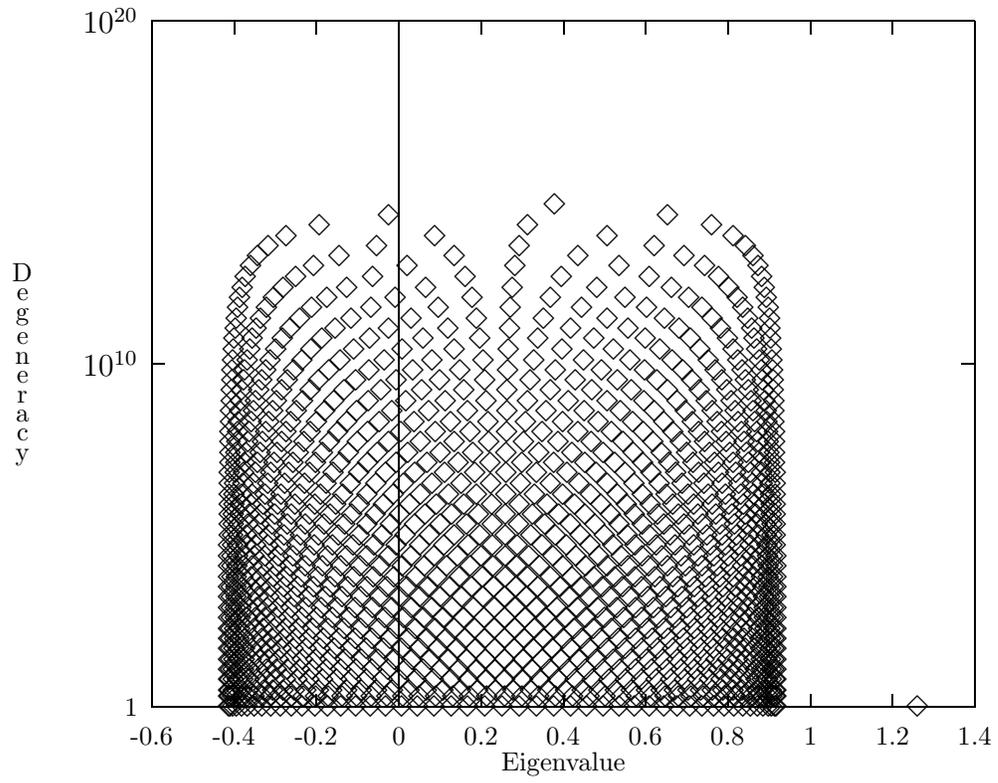
\item[Fig. 2]
Eigenvalues and their degeneracies for the synchronized state for $k=50$
generations.  The parameters are $h_p=.7$, $h_0=.2$, $h_s=.05$ and
$\lambda=1.26$. One can clearly see the gap that separates a nondegenerate
eigenvalue greater than unity and all the others below unity. Degeneracies
are on logarithmic scale for clarity.
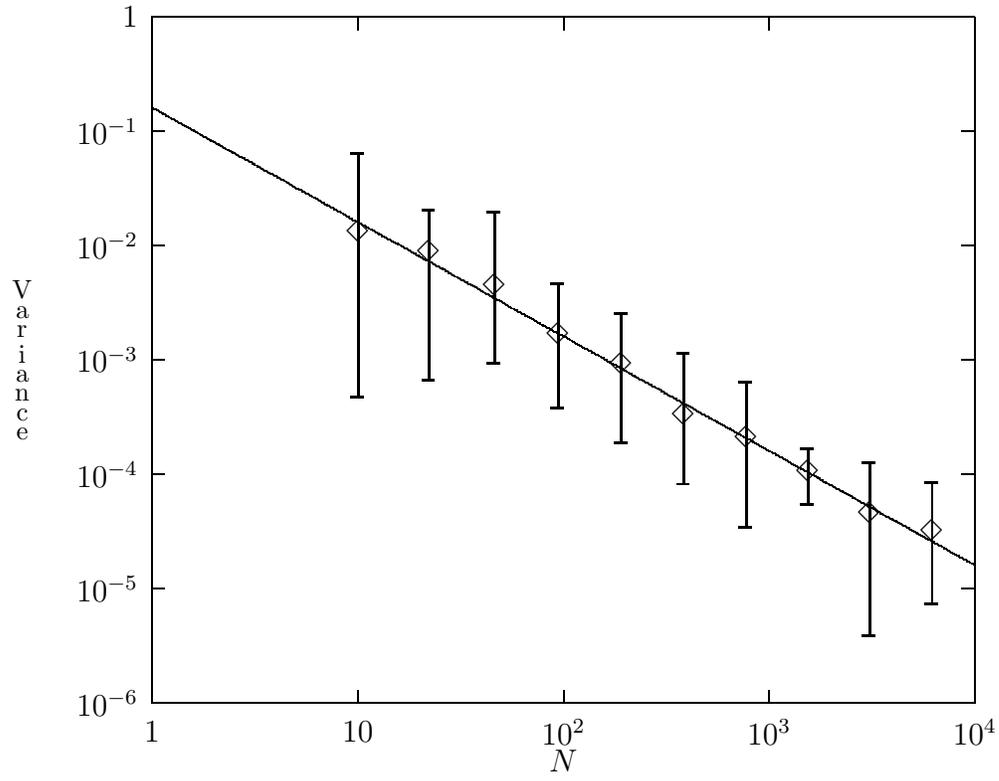
\item[Fig. 3]
Standard deviation of the mean field as a function of total number of
sites for the CML with function $f(x)=\mu x(1-x)$, with $\mu = 4.0$ and
$\epsilon = 0.1$. Similar behaviour also obtains for other values of $\mu,
\epsilon$.
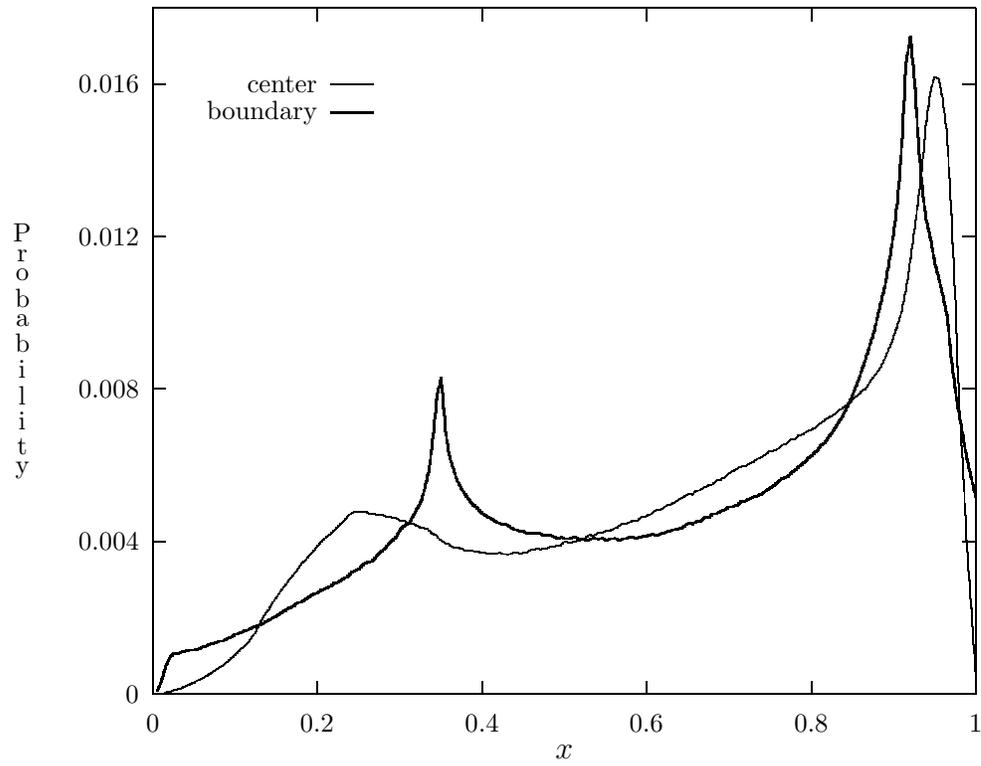
\item[Fig. 4]
Probability distributions of the CML at a central point and at the
boundary for the same set of parameters as in Fig. 3. The lattice has
$k=6$, and the first $10^4$ iterates are treated as transients. The
distribution is clearly very different for these two sites.
\end{itemize}
\newpage
\hspace*{\fill}

\begin{thebibliography}{99}
\bibitem{fn} The choice of $a$ in \cite{AGS} gives a larger
eigenvalue than that quoted, so we chose $ a = 1.5$ corresponding to the
eigenvalue $\approx$ 1.22. The evolution with map $1.26x
\;mod \;1$ also give similar results.  Note also that the extent
of synchronization depends on the noise level in the evolution.  Our linear
stability analysis merely guarantees the stability of the macroscopic
coherent pattern under infinitesimal perturbations in stationary frame.
\end{thebibliography}
\end{document}